\documentclass[prb,twocolumn,groupedaddress,showpacs]{revtex4}
\usepackage{graphicx} 
\begin{document} 
\bibliographystyle{prbrev}

\title{Condensation of helium in nanoscopic alkali wedges at zero temperature}

\author{E. S. Hern\'andez}
\affiliation{Departamento de F\'{\i}sica, Facultad de Ciencias Exactas
y Naturales, Universidad de Buenos Aires, \\
and Consejo Nacional de Investigaciones Cient\'{\i}ficas y T\'ecnicas,
Argentina}

\author{F. Ancilotto}
\affiliation{
Dipartimento di Fisica `G. Galilei',
Universit\`a di Padova, via Marzolo 8, I-35131 Padova, Italy,
and DEMOCRITOS National Simulation Center, Trieste, Italy}

\author{M. Barranco}
\affiliation{Departament ECM,
Facultat de F\'{\i}sica, Universitat de Barcelona, 08028 Barcelona, Spain}

\author{R. Mayol}
\affiliation{Departament ECM,
Facultat de F\'{\i}sica, Universitat de Barcelona, 08028 Barcelona, Spain}

\author{M. Pi}
\affiliation{Departament ECM,
Facultat de F\'{\i}sica, Universitat de Barcelona, 08028 Barcelona, Spain}
\begin{abstract}

We present a complete calculation of the  structure 
of liquid $^4$He confined to a concave nanoscopic wedge, as a
function of the opening angle of the walls. This is achieved
within a finite-range density functional formalism. The results here 
presented, restricted to alkali metal substrates, illustrate the change 
in meniscus shape from
rather broad to narrow wedges on  weak  and strong alkali
adsorbers, and relate this change to the wetting behavior of helium on
the corresponding planar substrate. 
As the wedge angle is varied, we find a sequence of stable states
that, in the case of cesium, undergo one filling and one emptying
transition at large and small openings, respectively. 
A computationally unambiguous criterion to determine the contact angle of 
$^4$He on cesium is also proposed.

\end{abstract}


\date{\today}
\maketitle

\section{Introduction}

The current possibility of tailoring the production of 
adsorbing surfaces in the nanoscopic scale, raises a series of questions
on the role of inhomogeneities in the 
growth of liquid films on nanopatterned planar
substrates. In particular, the wedge geometry is 
a simple one that permitted to carry on many theoretical anticipations
of the wetting behavior of fluids in such confinement (see,
e.g., Refs. \onlinecite{rascon00a,rascon00b,henderson04} and cited therein),
and it has been  recently argued\cite{henderson04} 
that models of fluids adsorbed in planar wedges can be 
regarded as a generic class of structured inhomogeneous fluids. 
 
Since
the early calculation of physisorption  of noble gases in an oblique
corner,\cite{cheng90} most
predictions of the shape changes undergone by the liquid meniscus
are based on the examination of the equilibrium liquid-vapor
interface  of a macroscopic sample at constant bulk density
$\rho_0$, whose sharp free surface is
described by a function   determined by
minimization of the surface free energy. The latter contains the energy cost
for building a curved liquid-vapor interface with surface tension 
$\sigma_{lv}$, in addition to the interfacial
contributions governed by  the surface energies 
$\sigma_{sl}$ and $\sigma_{sv}$, with $l, s, v$ denoting liquid,
solid and vapor/vacuum, respectively. These three surface tensions determine 
the contact angle $\theta$    as given by the classical 
  Young's relation.\cite{napior92,concus97,rejmer99}  These calculations confirm the
 thermodynamic theory of wetting in a wedge\cite{hauge92} that shows,
 on the ground of purely macroscopic arguments, that a filling
 transition characterized by a perfectly planar liquid-vapor interface
  takes place, at liquid-vapor coexistence, when the opening angle $2 \varphi$ 
of the walls
 is related to the contact angle  for a planar surface as
 $\varphi = \pi / 2 - \theta$. Moreover, the above theories classify
 the meniscus shapes for nonwetting liquids according
 to the following sequence:\cite{concus97, rejmer99}
\begin{itemize}
\item{
 $\theta > \frac{\displaystyle \pi}{\displaystyle 2} + \varphi$, 
spherical bridge;}
\item{ $\frac{\displaystyle \pi}{\displaystyle 2} + \varphi \ge \theta > 
\frac{\displaystyle \pi}{\displaystyle 2} - \varphi$,
  spherical convex meniscus;}
\item{ $\theta =  \frac{\displaystyle \pi}{\displaystyle 2} - \varphi$, 
planar meniscus;}
\item{$\theta <  \frac{\displaystyle \pi}{\displaystyle 2} - \varphi$, 
spherical concave meniscus.}
\end{itemize}
 
While classical, macroscopic approaches provide a solid frame for the
expected behavior of fluids in wedges, one may wonder if these
predicted phenomena are robust at the nanoscale. 
A study of interfacial phenomena in power-law wedges permits
 to trace the route from wetting to capillary 
 condensation\cite{rascon00b}  and suggests the possibility of
 adapting the  adsorptive abilities of solid substrates by shaping its
 surface inhomogeneities at the nanoscopic level. Recent measurements
 of the growth of Ar films on an array of
 microscopic linear wedges\cite{bruschi02} confirm this conjecture and
demonstrate  a clear crossover between a
  planarlike and a geometry-dependent behavior. 
On the theoretical side, it has
been  remarked\cite{rejmer99} that the full one-body density
$\rho({\bf r})$ for the confined particles should be obtained either from 
simulations or from density functional (DF) theory.  With the latter 
instrument,  we have recently investigated the growth of nanoscopic $^4$He
 clusters on planar alkali surfaces at zero temperature ($T$),
 employing a finite range DF.\cite{barranco03} 
The analysis of the energetics and shape systematics
with increasing number of atoms $N$ shows that 
helium drops on wettable alkali surfaces reach a
 maximum vertical height equal to the thickness of the prewetting
 film,  with the chemical potential at the prewetting jump in the zero 
temperature adsorption isotherm. These results illustrate that
 DF-based calculations can account for  the interplay between 
geometry and thermodynamics in nanoscopic systems,  sufficiently large 
to  remain beyond the reach of present microscopic many-body theories. 
On  these theoretical grounds, in this work we carry out a systematic 
investigation of the spatial structure and shape evolution of  nanoscopic 
helium samples in linear alkali wedges, as described in the next
Sections. In particular, in Sec. \ref{sec:method} we shortly review the
DF formalism employed, and in Sec. \ref{sec:results} we present our results for
 Cs and Na wedges, which are summarized in Sec. \ref{sec:summary}.

\section{Method}
\label{sec:method}

Our geometry is translationally invariant along one
spatial direction (y-axis). The equilibrium $^4$He density 
profile $\rho(x, z)$ will thus
depend on the $(x,z)$ coordinates only.
We compute the density profile that minimizes
 the zero temperature
grandpotential $\Omega= E - \mu \,N$, with $\mu$ the chemical potential of 
the helium atoms and with $N$ the grandcanonical ensemble average of 
the particle number operator, that  
in a wedge  extending a macroscopic length $L$ along the $y$-direction, takes 
the form
\begin{equation}
N = \int  d^3 {\bf r}\,\rho({\bf r}) \equiv L\, \int \int d x\, d
z\,\rho(x, z)   \; .
\label{eq2}
\end{equation}
 The variation $\delta \Omega/\delta \rho$ 
gives rise to the integrodifferential Euler-Lagrange equation
\begin{equation}
\left[-\frac{\hbar^2}{2 m_4}\,\nabla^2  +
V\left(\rho\right)+V_s(x, z)\right]\,\sqrt{\rho(x, z)}
 =\mu\,\sqrt{\rho(x, z)}
\label{eq1}
\end{equation}
Here $V(\rho)$ is the effective potential
arising from functional differentiation of the 
potential energy density per unit length, and $V_s(x, z)$ is 
the wedge potential. 
In this work, we have selected
the full Orsay-Trento (OT) DF,\cite{dalfovo95}
which is known to provide a quite accurate description of
inhomogeneous structures of $^4$He at $T=0$.
We can safely neglect gravitational effects in our calculations,
since the capillary length $a=[2\sigma_ {lv}/(g m_4\rho _0)]^{1/2}$, with 
$\rho _0 = 0.0218\, \AA ^{-3}$ the saturation density of $^4$He and $g$ the 
acceleration of gravity, is much larger than any linear dimension of the 
$^4$He samples considered in the present work.

\begin{figure}
\centerline{\includegraphics[width=0.9\columnwidth,clip]{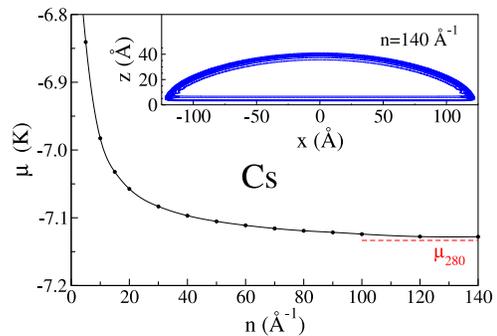}}
\caption{(Color online)
Chemical potential of a $^4$He pancake on planar Cs
as a function of linear density.
The chemical potential of the $n=280$ \AA$^{-1}$ pancake is indicated by
an horizontal dashed line.
The inset shows, for the $n=140$ \AA$^{-1}$ density profile,
the equidensity lines from $\rho=0.9 \rho_0$
to $0.1 \rho_0$ in $0.1 \rho_0$ steps.
 }
\label{fig1}
\end{figure}

To our knowledge, the best available adsorbing potential  for helium on
alkali planar surfaces is the Chizmeshya-Cole-Zaremba (CCZ) 
potential.\cite{chizmeshya98}
As a compromise between rigor and
physical insight,\cite{cole05} in the present work we have chosen to
approximate $V_s(x,z)$ by the summation of two planar
CCZ potentials at an angle $2 \varphi$. Double-counting effects, even
when sizeable, are only important for fairly large values of $2\varphi$ and 
in the vicinity of the wedge apex.\cite{henderson04}
Thus,  one should not expect qualitative
differences in the density patterns  for the fairly large
linear densities  considered here.

Equation (\ref{eq1}) is discretized  with 13-point formulas for the partial
derivatives and solved by an imaginary time method,\cite{barranco03}
employing a Fast Fourier Transform algorithm\cite{FFT} to obtain
$V(\rho)$ from the atom density $\rho(x, z)$.\cite{ancilotto98}
We have verified that the numerical outcome is  stable against the increase 
of the number of mesh points and the order of the discretization formulas. 

\section{Results}
\label{sec:results}

\subsection{Cesium wedges}

We address first the most interesting case of cesium, which is not wetted 
by $^4$He at $T=0$.
For reference, we have computed  a sequence of $^4$He pancake-shaped
systems on a flat Cs surface, namely,
 translationally invariant systems in the $y$-direction and
 characterized by the linear
density $n=N/L$. Fig. \ref{fig1} shows the chemical potential of the
$^4$He atoms  as a function of $n$, that displays a
neat tendency towards saturation at the bulk figure 
$\mu_0=-7.15$ K. The equidensity
lines for the density profile corresponding to a linear density
 $n$ = 140 \AA$^{-1}$ are shown in the inset of this figure. The profiles are
qualitatively similar to those shown  in Ref. \onlinecite{barranco03}  
in the case of the deposited drops, and confirm the
nanoscopic spatial scale of the systems. 

It should be kept in mind that in a typical adsorption experiment, one 
changes the chemical potential by varying the pressure in the chamber 
towards saturated vapor pressure (SVP) conditions.
Very close to SVP (but below it) the thermodynamic equilibrium
state of liquid $^4$He on Cs is a thin, microscopic planar film
partially wetting the surface. 
Nonetheless macroscopic, metastable $^4$He droplets can  be realized
if a finite amount of helium is deposited on the Cs surface, as shown 
 for instance in the experimental results of
Ref. \onlinecite{ross97}. The contact angle of such droplets
with the surface plane is a meaningful quantity to be measured/calculated 
and which can be related to thermodynamic -equilibrium- properties.
We shall see in the following that 
similar non-equilibrium configurations are found in the wedge geometry,
which might be experimentally realized  under particular conditions.

\begin{figure}
\centerline{\includegraphics[width=0.9\columnwidth,clip]{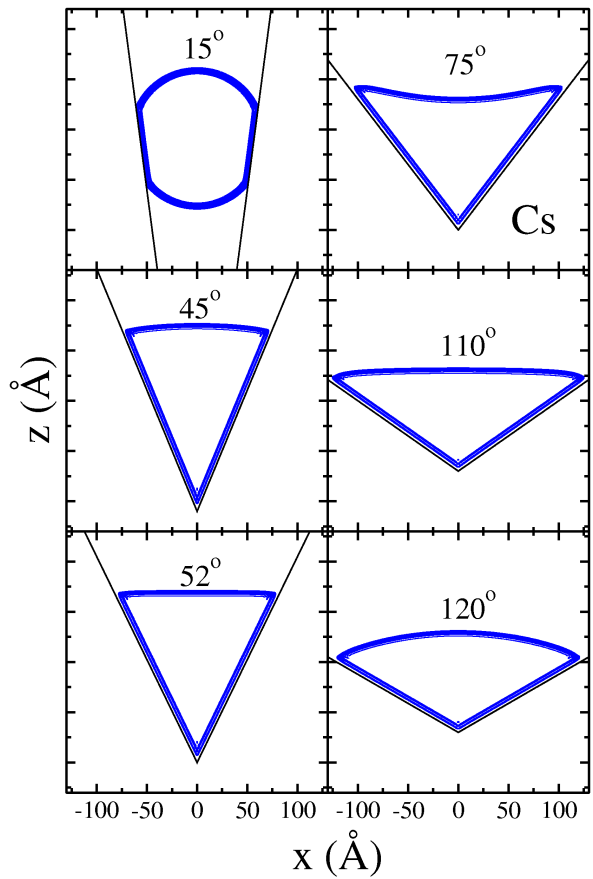}}
\caption{ (Color online)
Density profiles of helium at $n=280$
\AA$^{-1}$ in a Cs corner represented by thin straight lines,
for angular opening $2\varphi$ = 15, 45, 52, 75, 110,
and 120$^o$.
The distance between major tick marks on the $z$ axis is
50 \AA. The center of the 15$^o$ configuration
is about 400 \AA $\;$ above the apex of the wedge.
Equidensity lines are drawn as in Fig. \ref{fig1}.
 }
\label{fig2}
\end{figure}  
  
When  helium at given
$n$ is placed in a Cs wedge and  the opening angle $2\varphi$
is varied, we
generate the shape sequence illustrated in Fig. \ref{fig2}, at a linear
density
of 280 \AA$^{-1}$. 
 This evolution illustrates the modifications of the wetting
behavior of helium on Cs due to  the confinement
exerted by the corner. A convex helium pancake 
like the one shown in the inset of Fig. \ref{fig1} initially
adsorbed on a flat Cs surface, 
when subjected to the formation of a
wedge and decrease of the opening angle experiences a sequence of
changes
whose highlights are: a) a filling transition where the 
curvature vanishes, for an opening near 110$^o$, 
b) formation of a concave meniscus
that persists within a sizable range of angles;
 c) an `emptying' transition where a flat meniscus appears near 52$^o$,
with
restoration of the convex  shape for smaller angles; d) formation of a 
bridge that will be expelled from the wedge for a sufficiently small angle. 
We have verified that the bridge profiles are perfectly
circular, as predicted by the classical 
theories.\cite{cheng90,napior92,concus97,rejmer99} Note, however, that here 
the term `transition' does not address a thermodynamic phase
transition, but
 a crossover between two different, well defined regimes characterized by
 geometry and energetics as analyzed in what follows.
 
\begin{figure}
\centerline{\includegraphics[width=0.9\columnwidth,clip]{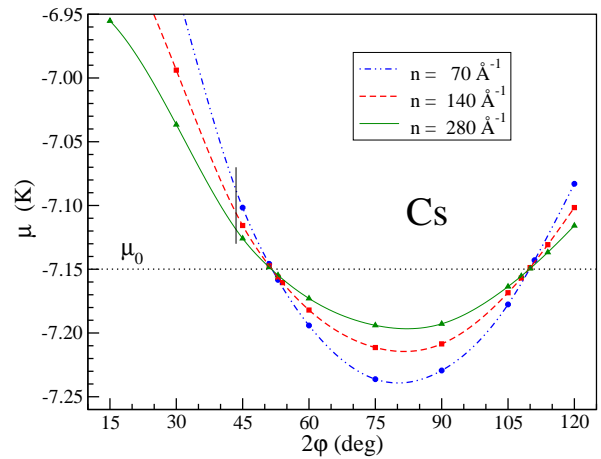}}
\caption{(Color online) 
Chemical potential of helium atoms in a Cs wedge as a function of angular opening
$2 \varphi$ (degrees) for linear densities 70, 140 and 280 \AA$^{-1}$.
Wedge configurations to the left of the thin vertical segment at about
 43$^o$ have higher energies than a pancake,
with the same linear density, adsorbed on a planar surface.
}
\label{fig3}
\end{figure}

\begin{figure}
\centerline{\includegraphics[width=0.9\columnwidth,clip]{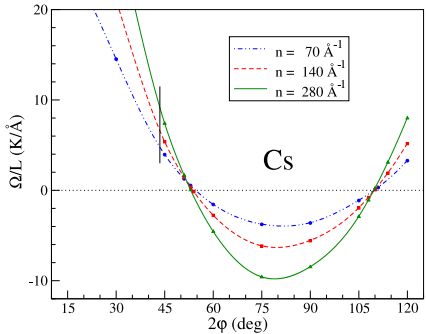}}
\caption{(Color online)
Same as Fig. \ref{fig3} for the grandpotential per unit length.
}
\label{fig4}
\end{figure}

The above sequence can be 
understood by examining  the energetics
of helium atoms at several  linear densities, as
functions of the wedge  opening. For this sake, 
in Fig. \ref{fig3} we plot the chemical potential for $n$ = 70, 140 and
280 \AA$^{-1}$. Several interesting features become evident. First, the
three curves cross at the bulk value $\mu_0$ for two opening angles, namely 
the filling one  2$\varphi_F$ = 110$^o$  and the emptying one
2$\varphi_E$ = 52$^o$. Within this interval,  the
helium chemical potential $\mu$ lies below the thermodynamic limit $\mu_0$, 
and the systems are thermodynamically stable, since $d \mu/ d n$ is positive. 
By contrast, above the filling angle $\varphi_F$ and below the emptying
angle $\varphi_E$, $\mu$ lies above the bulk value with negative
$ d \mu/ d n$; the  corresponding helium samples are then 
metastable. These characteristics can be further analyzed 
looking at the grandpotential per unit length
$\Omega/L = E/L - \mu n$, where $E/L$ is the energy per unit length,
displayed in  Fig. \ref{fig4} for the same linear densities.
Note that in the present context, $\Omega/L$ represents the 
surface-plus-line energy difference per unit length between a wedge
filled with liquid and an empty wedge.\cite{rejmer99} At
$\varphi_F$, $\Omega/L$ is independent of $n$ and vanishes for
these large, however nanoscopic, amounts of fluid. 
Thus, the conditions for a filling transition, 
previously derived for 
macroscopic systems,\cite{rejmer99} are fulfilled for these
helium samples.  This  suggests an effective way of determining the contact 
angle $\theta$ of  $^4$He
on Cs by means of DF calculations; we recall that previous DF--based
calculations
 of this angle give values between 31$^o$
(Ref. \onlinecite{ancilotto00}) and 36$^o$,
\cite{ancilotto98} in good agreement with experiments in Refs.
\onlinecite{rolley97,ross97}.
Other DF estimates provide somewhat smaller
values,\cite{szybisz03} whereas a different experimental determination
of the contact angle\cite{Kli95} yields $\theta = 48^o \pm 1^o$.
Using the value of $\varphi_F\sim 110^o$ where the curves cross in Fig. \ref{fig4}  
we obtain here  a contact angle $\theta = \pi/2 - \varphi_F$ = 35$^o$.

\begin{figure}
\centerline{\includegraphics[width=0.9\columnwidth,clip]{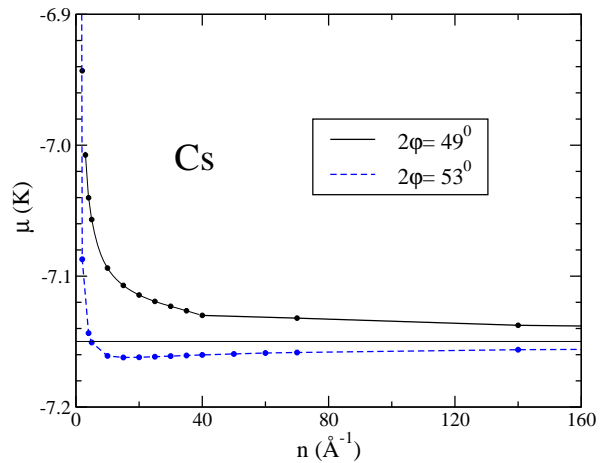}}
\caption{ (Color online)
Chemical potential of helium atoms in a Cs wedge
as a function of linear densities.
Solid line: 2$\varphi =49^o$; dashed line: 2$\varphi =53^o$.
}
\label{fig5}
\end{figure}

Figure \ref{fig4} also shows, in correspondence with Fig. \ref{fig3}, 
that a second curve crossing, where $\Omega/L$ is
$n$-independent, occurs at the aperture $2\varphi_E \sim 53^o$
i.e., at the threshold for the formation of the
bridge, which we have found to be clearly visible only
below $\sim 30^o$.
Except in a very narrow angular region near
$\varphi_E$, whose existence is likely
attributable to finite size effects,\cite{note} $\Omega/L$
is negative in the interval $[\varphi_E,\varphi_F]$,
indicating that condensation might occur either in
a continuous way or 
through discontinuous jumps in the linear coverage. 
In fact, the latter is the case here, as shown below in relation to Fig. 
\ref{fig5}.

Within the interval $[\varphi_E,\varphi_F]$, the gain in energy originates in 
the disappearance of the solid-vacuum interface at the meniscus
contact line, that leaves only a liquid-vacuum interface
at the free surface, as seen in Fig. \ref{fig2}. 
This is reversed for $\varphi < \varphi_E$ 
since the system shifts to a nonwetting regime.
In these wedges, the energy balance favors the
 presence of vacuum, rather than liquid, at the corner,
making room to the formation of the inner convex meniscus that
`dries' the corner and forms the bridge,
at the price of  increasing the total energy.

The existence of two stable solutions in the $[\varphi_E,\varphi_F]$ 
interval for a given coverage as seen in Fig. \ref{fig3}, and in
flat meniscus configurations at $\mu=\mu_0$,
is a genuine nanoscopic effect, as it does not appear for
macroscopic samples,
where only a filling transition is predicted for liquid
in a wedge made of partially wettable surfaces.\cite{rejmer99} 
As a matter of fact, {\em all}
equilibrium macroscopic configurations, irrespective of the opening,
might be thought of
as corresponding to the {\em same} chemical potential, that of 
bulk liquid, with the fluid density in the wedge taking its value at
bulk liquid-vacuum coexistence at zero temperature.

To illustrate the mechanism of condensation, in Fig. \ref{fig5}  
we plot two selected adsorption isotherms $\mu (n)$, for angles $2\varphi 
= 49$ and $53^o$. The respective $n=0$ ordinates -not shown in the
current scale- correspond to the binding energy of a single helium atom to the 
wedge at each aperture, computed separately. The solid line ($2\varphi =49^o$) 
represents $\mu (n)$ for an opening slightly below $\varphi _E$: the 
$^4$He configurations in such wedge are always metastable with respect 
to the bulk, since $\mu (n)$ approaches $\mu _0$ from above. 
Thus, no condensation can occur in the wedge at such an aperture. 
By contrast, when the angle is slightly larger that 
$\varphi _E$ (dashed line, $2\varphi =53^o$), 
condensation takes place with a jump in the linear density from 
$n=0$ to a finite value. The associated `pre-emptying'
transitions occur at a value of the chemical 
potential which can be found by using the usual equal-area Maxwell construction
(or, equivalently, by identifying the crossing between the $\mu (n) $ and
$E/N$ curves.\cite{barranco03}) From this construction we find that at 
$2\varphi =53^o$ a jump from $n=0$ to $n\sim 270\,\AA ^{-1}$ occurs 
at $\mu =-7.155\,$K, i.e. just below  SVP. 
This indicates that a crossover between a filled and an empty wedge is only 
meaningful for linear densities near to and above $\sim 270$ $\rm \AA^{-1}$,
i.e., more dilute helium vapors cannot condense in the wedge at angles equal to and 
slightly above 2$\varphi_E$.
A similar behavior (i.e. finite jumps in coverage below SVP) is 
encountered for all angles in the range $[2\varphi _E, 2\varphi _F]$; 
however, for the largest openings in this interval, condensation  occurs 
at rather low coverages, say $n \approx$ 5-10 \AA$^{-1}$.

The occurrence of an `emptying' regime at lower opening angle,
as suggested by our results, is surprising at first sight. 
No `emptying' transition
is predicted from macroscopic thermodynamic arguments,
and this behavior seems also to contradict the expectation that a 
narrow wedge should be filled by capillary condensation (CC), below SVP,
in the same way as small pores/slits are known, from theoretical
calculations, to undergo a CC transition even for weakly adsorbing 
surfaces.\cite{Gat99,Szy02} We believe that the wedge geometry makes 
here an important difference with respect to pore/slit case
-even in the narrowest wedge, the distance between the walls can 
be arbitrarily large sufficiently far from the apex.

The analysis of the energy per particle for fixed $n$, as a function of
wedge opening, indicates that for the linear densities under consideration,
below 2$\varphi \approx$ 43$^o$  helium bridges are metastable 
(i.e. have higher energies) with respect to helium pancakes 
deposited on planar Cs surfaces.
This limit is shown in Fig. \ref{fig4} by a vertical segment.
Similarly, wedges corresponding to openings 2$\varphi \gtrsim 120^o$ are
also metastable with respect to pancakes with the same number
of atoms. It should be remarked that these metastable configurations appear 
for finite average numbers of $^4$He atoms (per unit length) within a
grandcanonical description; as a reference, we recall that a droplet with 
finite atom number $N$ on Cs is metastable, not only with respect to bulk, but
 also  with respect to a thin film at finite coverage $N/A$.  In view of these
 results, we are 
confident that in an hypothetical experimental setup where {\it finite}, 
nanoscopic amount of He are deposited in Cs alkali wedges, our 
prediction that narrow wegdes  are either empty or hosting `bridge'
configurations could in principle be checked.

\subsection{Sodium wedges}

\begin{figure}[h!]
\centerline{\includegraphics[width=0.8\columnwidth,clip]{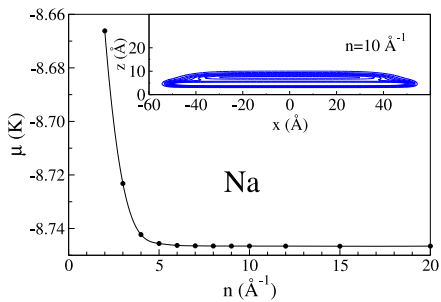}}
\caption{(Color online) Same as Fig. \ref{fig1} for $^4$He on planar Na.
In this case, the inset corresponds to $n=10$ \AA$^{-1}$.
 }
\label{fig6}
\end{figure}

We have carried out similar calculations
for helium on sodium, analyzing the evolution of helium 
initially
placed in the unstable regime at low coverages --i.e., below the 
prewetting  transition--. Fig. \ref{fig6} displays the 
chemical potential of $^4$He pancakes on  a flat Na surface.
For this strong adsorber, the corresponding
asymptotic limit is $-8.75$ K,  the value of the chemical
potential at the prewetting jump.\cite{barranco03}

\begin{figure}
\centerline{\includegraphics[width=0.9\columnwidth,clip]{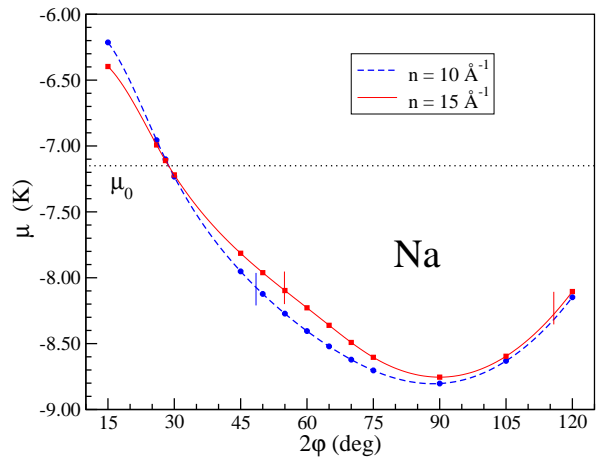}}
\caption{(Color online) 
Chemical potential of helium atoms in a Na wedge as a function of angular opening
$2 \varphi$ (degrees) for a linear densities $n=10$ and 15 \AA$^{-1}$.
For $n=15$ \AA$^{-1}$,  wedge configurations within the thin
vertical segments are stable with respect to the infinite
pancake configuration. 
For $n=10$ \AA$^{-1}$,  wedge configurations to the right  of the
corresponding thin vertical segment are stable with respect to the infinite
pancake configuration. 
 }
\label{fig7}
\end{figure}

\begin{figure}
\centerline{\includegraphics[width=0.9\columnwidth,clip]{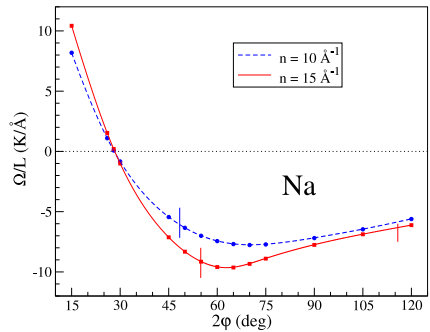}}
\caption{(Color online) 
Same as Fig. \ref{fig7} for the grandpotential per unit length.
}
\label{fig8}
\end{figure}

Figs. \ref{fig7} and \ref{fig8} display, respectively, the chemical potential
and grandpotential per unit length of a $^4$He sample with $n=15$ \AA$^{-1}$ 
and shows that the filling transition
is obviously absent due to the fact that sodium is wetted by $^4$He at
$T=0$.
In these figures, the thresholds for the stability of pancake structures
are shown by the short vertical segments.
 
\begin{figure}
\centerline{\includegraphics[width=0.9\columnwidth,clip]{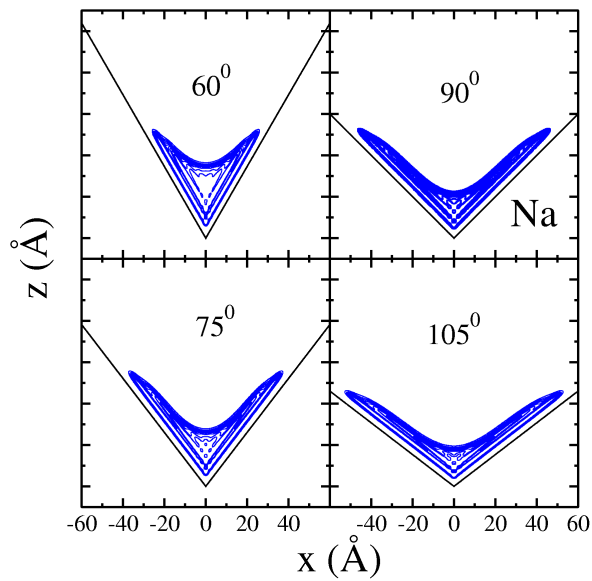}}
\caption{ (Color online)
Density profiles of helium at $n=15$
\AA$^{-1}$ in a Na corner represented by thin straight lines,
for angular opening $2\varphi$ = 60, 90, 105, and 120$^o$.
The distance between major tick marks on the $z$ axis is
20 \AA. 
Equidensity lines as in Fig. \ref{fig2}.
 }
\label{fig9}
\end{figure}

Finally, typical density patterns are displayed in
Fig. \ref{fig9} for the linear density $n=15$ \AA$^{-1}$ and
several wedge openings. The panels show the spread of the
wetting liquid, that retains a rather deep concave meniscus.
This figure shows that helium originally deposited on a flat,
strongly wettable adsorber like Na, just follows the folding of the
surface with angle $2 \varphi$ and forms a deep meniscus. 
For sufficiently small angles
the pattern evolves into a nanoscopic bridge,
similar to that shown in the first panel of Fig. \ref{fig2}.
Such configurations have, however, higher energies than the 
corresponding pancake structures with the same linear 
density, deposited on one of the walls,
and thus should not occur in sodium or any other wettable wedge.

\section{Summary}
\label{sec:summary}

We have presented the first detailed calculations of the
structure and energetics of nanoscopic samples of $^4$He in a
linear wedge with two identical alkali walls.
Our results indicate that for a Cs substrate, a sequence of states
that include one filling transition and a threshold for 
-metastable- configurations where the wedge is void of $^4$He
occurs by varying the wedge opening at
zero temperature, with condensation at negative grandpotential 
(accompanied by pre-filling jumps in the linear density) as an 
intermediate regime. When the  material is near saturation, the
angle for the filling transition is stable against changes in the linear
 density and can be identified with the contact angle on a
planar substrate. 
It is important to bear in mind that the experimental realization 
of such system, which could be done in practice by 
depositing the chosen alkali on a silicon surface previously
patterned with microscale wedges,\cite{mistura}  is very challenging
due to the reactive character of alkalis.

The present wedge geometry can be also used as a first step to model
rough alkali surfaces.
Quite surprisingly, our calculations  indicate that 
narrow Cs wedges (of nanoscopic size)
are not filled by finite amounts of helium. This circumstance might be 
relevant in
the study of the wetting behavior of $^4$He on rough Cs films, and in
the modelling of helium nanopuddles formed on strong-pinning
surfaces.\cite{Kli05}

We have also carried out similar calculations
for helium on sodium which indicate that
a helium pancake originally deposited on a flat,
strongly wettable adsorber like Na, basically follows the folding of the
surface with angle $2 \varphi$ and forms a deep meniscus. Lastly,
we have shown that a theoretical search of the vanishing of the
grandpotential as a function of wedge opening, along with liquid-vacuum 
coexistence at $\mu_0$ and independent on the linear coverage $n$, 
is feasible within a DF
description and unambiguously determines, from the theory point of view,
the
contact angle of a nanoscopic helium sample on a nonwettable substrate.

\acknowledgments

We are indebted to F. Caupin, M.W. Cole, R. Estalella, H. Godfrin, G.
Mistura, and A.F.G. Wyatt for stimulating discussions and  helpful advice.
This work was supported by grants PIP2391/00
from CONICET and PICT03-08450 from ANPCYT, Argentina,  FIS2005-01414
from DGI, Spain (FEDER), and 2005SGR-00343 from Generalitat of
Catalunya. F. A. acknowledges funding from CESCA-CEPBA, Barcelona, through 
the program HPC-Europa Transnational Access, and Project num. CPDA033545 
of Padua University.


\begin{thebibliography}{99}
\bibitem{note} We recall that in the macroscopic limit, the
conditions $\Omega/L=0$, $\Omega/L$ independent of $n$, and
$\mu=\mu_0$, happen simultaneously.
\end{thebibliography}
\end{document}